# Three Barriers to Kantian Cooperation under Inequality

Igor Sloev, School of Computational Social Sciences, European University at St. Petersburg, Russia, 191187, St. Petersburg , 6/1A Gagarinskaya st., St. Petersburg 191187 Russia. PhD, Associated Professor, sia@eu.spb.ru, +7-911-199-43-65.

Gerasimos Lianos, Higher Colleges of Technology, Abu Dhabi College, Baniyas, Abu Dhabi, UAE, PhD, Associated Professor, glianos@hct.ac.ae, +971-50-430-92-66

## Abstract

Multiplicative Kantian equilibrium, introduced by Roemer, has been proposed as a solution to inefficiency in social dilemmas, yet its emergence and stability are associated with a number of coordination and distributional difficulties. Drawing on a simple model of voluntary public good provision with two agents who differ in their initial endowments and have logarithmic preferences, this paper identifies three barriers that arise sequentially when unequal agents attempt to adopt Kantian cooperation voluntarily. The model builds on the classical setup of Bergstrom, Blume, and Varian and compares Nash equilibrium with multiplicative Kantian equilibrium under two different parametrizations: one in which the strategic variable is contributions to the public good, and another in which it is private consumption. First, we show that the transition from Nash equilibrium to contribution-space Kantian behavior is not always a Pareto improvement: under sufficiently high inequality, the poor agent may prefer the Nash outcome. This barrier can be mitigated by preliminary redistribution. Second, even when agents are willing to cooperate, the choice of a parametrization of the strategic space becomes a distributional issue: different ways of scaling actions lead to different distributive consequences and create a conflict of interest between agents. Third, if the chosen parametrization admits a continuum of Pareto-efficient outcomes, an additional coordination problem arises—agreeing on a specific point on the Pareto frontier. The paper reconstructs these barriers as a three-step coordination problem in which expectations about later stages affect agents' willingness to enter Kantian cooperation at the outset. On the basis of the results obtained, a program for further formal research is outlined. The findings contribute to understanding the conditions under which Kantian cooperation can be voluntarily adopted and sustainably maintained.

## 1. Introduction

Standard noncooperative game theory predicts that in social dilemmas—voluntary provision of public goods, use of common-pool resources, emission reductions—individually rational behavior leads to collectively inefficient outcomes [Hardin, 1968; Bergstrom, Blume, Varian, 1986; Ostrom, 1990]. The concept of multiplicative Kantian equilibrium (hereafter MKE), introduced by J. Roemer [Roemer, 2010; Roemer, 2019], represents an attempt to overcome this inefficiency. In contrast to Nash equilibrium, an agent in MKE chooses an action by mentally adjusting her own strategy in the same proportion as she would wish all other agents to adjust theirs. This behavioral norm, interpretable as a formalization of the Kantian categorical imperative, can yield Pareto-efficient outcomes under unidirectional externalities.

The MKE concept has been developed in a wide range of applied studies: from tax competition [Eichner, Pethig, 2020] and the dynamics of climate agreements [Grafton, Kompas, Long, 2017] to corporate environmentalism [Kurata, Long, 2025], the analysis of R&D and oligopoly [Donduran, Ünveren, 2021], vaccination [De Donder et al., 2025], and international cooperation [Ulph, Ulph, 2024]. Yet, despite this breadth of applications, three fundamental questions remain insufficiently clarified: under what conditions agents switch to Kantian behavior in the first place; which specific form of Kantianism they select when alternative parametrizations are feasible; and how they can coordinate on a particular efficient outcome when the set of such outcomes is not a singleton. The present paper examines how these three challenges affect the very possibility of a transition to stable Kantian cooperation.

We do not address the widely discussed commitment problem—the individual incentive to deviate from the Kantian norm when others comply with it. We assume that agents are capable of binding themselves in one way or another, be it through trust, repeated interaction, or institutional mechanisms. As we demonstrate below, even when the commitment problem is resolved, the transition to stable Kantian cooperation is not guaranteed: three additional barriers arise along the way.

The MKE concept was originally designed to accommodate heterogeneous agents [Roemer, 2019, ch. 3]; however, most applied works either assume symmetry or treat the strategy scale as exogenously given. The issue of choosing a parametrization has merely been outlined: Roemer [2019] stressed the importance of a coordinated choice of strategies, Sher [2020] pointed out the normative implications of strategic non-equivalence, and Roemer [2020] insisted that Kantianism itself presupposes symmetry. We go further and place heterogeneity at the center of the analysis. Using a simple model of a public-good economy with two agents who share identical logarithmic preferences but differ in initial endowments, we show that several natural parametrizations may exist and that agents' interests regarding the choice among them can be strictly opposed. The model builds on the classical setup of Bergstrom, Blume, and Varian [1986] and considers three equilibria: the standard Nash equilibrium, MKE in the space of contributions (b-MKE), and MKE in the space of private consumption (c-MKE).

The contribution of the paper is threefold. First, at the level of formal results, the model reveals three effects that have received little attention in the literature: (i) a threshold inequality condition beyond which Kantian cooperation ceases to be a Pareto improvement over Nash and cannot arise voluntarily; (ii) a conflict of interest between agents regarding the choice of parametrization, which shows that strategic non-equivalence is not a technical curiosity but a distributional issue; and (iii) a continuum of Pareto-efficient equilibria under the consumption-space MKE, which gives rise to an equilibrium-selection problem absent from the contribution-space counterpart. Preliminary redistribution is shown to be capable of removing the first barrier, and the selection problem is argued to echo the logic of the Second Welfare Theorem [Arrow, 1951; Debreu, 1959; Mas-Colell, Whinston, Green, 1995], yet has not previously been emphasized in the context of Kantian cooperation.

Second, at the conceptual level, we demonstrate that these three barriers are not isolated but form a logical sequence. Expectations about future decisions link them into a unified dynamic structure in which the entry threshold into Kantianism depends not only on initial inequality but also on the institutional capacity of a society to coordinate the subsequent steps.

Third, building on this analysis, we outline a research program consisting of formal questions that flow directly from the model and call for further investigation.

The paper is organized as follows. Section 2 introduces the model. Section 3 presents the main formal results. Section 4 provides a substantive interpretation, including a sequential reconstruction of the three barriers and an outline of the research program. Section 5 concludes.

## 2. Model

Consider an economy with two agents, $i = 1, 2$, each endowed with an initial stock of a private good $e_i > 0$. Agents' preferences are represented by the utility function

$$U_i(B, c_i) = \ln B + \ln c_i,$$

where $B = b_1 + b_2$ is the total amount of the public good, $b_i \geq 0$ is agent $i$'s voluntary contribution, and $c_i = e_i - b_i$ is his private consumption. This specification is a special case of the Bergstrom–Blume–Varian [1986] model and is widely used for analyzing voluntary provision of public goods.

We consider two equilibrium concepts—Nash and MKE—where MKE is analyzed under two different parametrizations: in the space of contributions (b-MKE) and in the space of private consumption (c-MKE). We focus on these two parametrizations because they correspond to two economically meaningful interpretations of Kantian symmetry: equality of sacrifice (b-MKE) and equality of outcome (c-MKE).

**Nash equilibrium (NE).** Each agent maximizes his own utility, taking the other's strategy as given. Suppose $e_1 \leq e_2$ (agent 1 is poor, agent 2 is rich). If $e_2 \leq 2e_1$, both agents contribute a positive amount. From the first-order conditions $dU_i/db_i = 1/B - 1/c_i = 0$ we obtain $c_i = B$. Summing the budget constraints $c_i + b_i = e_i$ yields $B = (e_1 + e_2)/3$. The equilibrium contributions and consumption levels are

$$b_i^{NE} = \left(2e_i - e_j\right)/3, c_i^{NE} = (e_1 + e_2)/3, i \neq j.$$

The equilibrium utilities are equal:

$$U_i^{NE} = 2\ln\big((e_1 + e_2)/3\big),$$

which illustrates the well-known neutrality theorem [Bergstrom, Blume, Varian, 1986].

If $e_2 > 2e_1$, the poorer agent contributes nothing. The Nash equilibrium is then

$$b_1^{NE} = 0, b_2^{NE} = e_2/2, c_1^{NE} = e_1, c_2^{NE} = e_2/2,$$

with utilities

$$U_1^{NE} = ln(e_2/2) + lne_1, U_2^{NE} = 2ln(e_2/2).$$

**MKE in the contribution space (b-MKE).** Suppose agents treat contributions $(b_1, b_2)$ as their strategic variables. Substituting the budget constraint $c_i = e_i - b_i$ into the original utility function yields the payoff as a function of contributions:

$$W_i^b(b_1, b_2) = \ln(b_1 + b_2) + \ln(e_i - b_i).$$

Following Roemer [2010], a strategy profile $(b_1, b_2)$ is an MKE if no agent can increase his utility by multiplying the entire contribution vector by a scalar $a \geq 0$ such that $e_i - ab_i \geq 0$. Under differentiability, this is equivalent to

$$(d/da)W_i^b(ab_1, ab_2)|_{a=1} = 0.$$

Differentiating and evaluating at $a = 1$ gives

$$b_i/(e_i - b_i) = 1 \Rightarrow b_i^{bMKE} = e_i/2.$$

This equilibrium is unique and, by Roemer's [2010] theorem, Pareto efficient. The utilities are

$$U_i^{bMKE} = \ln\big((e_1 + e_2)/2\big) + \ln(e_i/2) = \ln\big(e_i(e_1 + e_2)\big) - 2\ln 2.$$

**MKE in the private-consumption space (c-MKE).** Alternatively, agents may treat private consumption $c_i$ as the strategic variable. Expressing contributions from the budget constraint $b_i = e_i - c_i$ and substituting into the original utility gives the payoff as a function of consumption:

$$W_i^c(c_1, c_2) = \ln\big((e_1 - c_1) + (e_2 - c_2)\big) + \ln c_i.$$

The MKE condition for a profile $(c_1, c_2)$,

$$(d/da)W_i^c(ac_1, ac_2)|_{a=1} = 0,$$

leads to

$$(c_1 + c_2)/(e_1 + e_2 - c_1 - c_2) = 1 \Rightarrow c_1 + c_2 = (e_1 + e_2)/2.$$

The set of c-MKE is therefore a straight line segment, subject to individual feasibility constraints $0 < c_i \leq e_i$. All these outcomes are Pareto efficient. Among them, one can single out the symmetric equilibrium $c_1 = c_2 = (e_1 + e_2)/4$, which is feasible whenever $e_2 \leq 3e_1$. At this symmetric allocation, the utilities are

$$U_i^{cMKE} = 2\ln(e_1 + e_2) - 3\ln 2.$$

It is important to stress that b-MKE and c-MKE are not two distinct games, but two alternative descriptions of the same feasible set: the budget constraint $b_i + c_i = e_i$ provides a one-to-one mapping between the two representations. Behaviorally, however, they embody two different understandings of what agents universalize in their Kantian reasoning. In b-MKE, contributions (sacrifices) are scaled, and symmetry is achieved in the space of efforts: every agent contributes the same share of his income. In c-MKE, private consumption (the residual) is scaled, and symmetry is achieved in the space of outcomes: every agent retains the same level of private good. Both logics appeal to the Kantian principle of symmetry, but to different symmetries, which is precisely what generates the substantive conflict analyzed below.

## 3. Formal Results

**Proposition 1 (Threshold inequality condition).**

*The poor agent gains from switching to b-MKE compared with the Nash equilibrium if and only if $e_2 < (5/4)e_1$. For the rich agent, switching to b-MKE is always an improvement.*

*Proof.* Compare the utilities of the poor agent. If $e_2 \leq 2e_1$, the Nash equilibrium is interior and

$$U_1^{bMKE} - U_1^{NE} = ln\big(e_1/(e_1 + e_2)\big) + 2ln(3/2),$$

which is positive exactly when $e_2 < (5/4)e_1$. If $e_2 > 2e_1$, the Nash equilibrium is corner and

$$U_1^{bMKE} - U_1^{NE} = ln\big((e_1 + e_2)/(2e_2)\big) < 0,$$

so the poor agent again prefers Nash.

For the rich agent, in the interior case

$U_2^{bMKE} - U_2^{NE} = \ln\big(e_2/(e_1 + e_2)\big) + 2\ln(3/2) > 0$ whenever $e_2 \geq e_1$;

in the corner case

$$U_2^{bMKE} - U_2^{NE} = ln\big((e_1 + e_2)/e_2\big) > 0.$$

Thus the rich agent always gains from the switch. ■

*Hence, under high inequality b-MKE ceases to be a Pareto improvement over Nash. In the absence of compensatory transfers or additional coordination mechanisms, unanimous adoption of Kantian cooperation becomes impossible—this is the first barrier.*

**Proposition 2 (Conflict between b-MKE and the equal-consumption c-MKE benchmark).**

*Provided that the symmetric c-MKE allocation is feasible ($e_2 \leq 3e_1$), the poor agent prefers symmetric c-MKE to b-MKE, whereas the rich agent has the opposite preference.*

*Proof.* Compare the utilities of the poor agent in the two Kantian equilibria:

$U_1^{bMKE} - U_1^{cMKE} = ln\big(2e_1/(e_1+e_2)\big) < 0$ (since $e_1 < e_2$).

For the rich agent,

$U_2^{bMKE} - U_2^{cMKE} = ln\big(2e_2/(e_1+e_2)\big) > 0.$

Thus, the agents' interests regarding the choice of "scale" are strictly opposed. ■

The ranking of preferences is as follows. When $e_2 < (5/4)e_1$, for the poor agent $U^{cMKE} > U^{bMKE} > U^{NE}$, while for the rich agent $U^{bMKE} > U^{cMKE} > U^{NE}$. When $e_2 > (5/4)e_1$, the poor agent ranks $U^{cMKE} > U^{NE} > U^{bMKE}$, whereas the rich agent's ranking remains $U^{bMKE} > U^{cMKE} > U^{NE}$.

In other words, strategic non-equivalence generates a distributive conflict: the choice of parametrization becomes not merely a technical issue but a distributional and institutional choice—the second barrier.

**Proposition 3 (Continuum of equilibria in c-MKE).**

*The set of c-MKE constitutes a one-parameter family of Pareto-efficient feasible outcomes. Parameterizing it by the poor agent's consumption share $\lambda = c_1/(c_1+c_2)$, the poor agent's utility increases in lambda while the rich agent's utility decreases.*

*Proof.* From $c_1 + c_2 = (e_1+e_2)/2$ it follows that $c_1$ can take any value in the interval $(0, mine_1, (e_1+e_2)/2]$. Expressing $c_2$ through $c_1$ and substituting into the utility functions yields

$U_1^{cMKE} = \ln\big((e_1+e_2)/2\big) + \ln c_1,$

$U_2^{cMKE} = \ln\big((e_1+e_2)/2\big) + \ln\big((e_1+e_2)/2 - c_1\big).$

The monotonicity in $\lambda = 2\mathrm{c}_1/(\mathrm{e}_1+\mathrm{e}_2)$ is obvious. ■

*Consequently, c-MKE raises the equilibrium-selection problem—the third barrier, which is absent in b-MKE. Proposition 2 relies on the symmetric c-MKE benchmark. Proposition 3 shows that once the symmetric benchmark is abandoned, the parametrization problem and the equilibrium-selection problem become intertwined: agents may disagree not only about the choice of parametrization but also about the particular allocation to be implemented within c-MKE.*

Thus, the three formal results reveal a logical chain: inequality blocks the entry into cooperation; the choice of parametrization generates a conflict of interest; and under a particular parametrization the equilibrium-selection problem arises. Taken together, these results show that the strategic non-equivalence of MKE is not merely a technical curiosity but a source of a fundamental conflict between two different understandings of Kantian fairness: solidarity in burden (b-MKE) and solidarity in outcome (c-MKE).

## 4. Discussion: Interpreting the Formal Results

### 4.1. The Entry Problem: Inequality as a Barrier

Proposition 1 shows that the switch to Kantian behavior is not automatic. If cooperation is by default realized in the contribution space (b-MKE), substantial inequality makes it unacceptable to the poorer participant. This finding, obtained for a specific functional form, raises a general question: at what level of inequality and for which class of preferences can Kantian cooperation be blocked at the outset? The result echoes the idea of the "veil of ignorance" [Rawls, 1971], which could lead agents to choose a more equitable parametrization that guarantees an improvement for the least advantaged. The model also allows us to pose the question of the role of redistribution: if the rich agent compensates the poor one ex ante for the losses from switching to Kantianism, would this be beneficial for both sides?

The following example illustrates that voluntary redistribution may remove the entry barrier. Consider the case where inequality only slightly exceeds the threshold: $e_2 = (5/4)e_1 + \delta$, with $\delta > 0$ small. Then, by Proposition 1, the poor agent strictly prefers the Nash outcome to the Kantian outcome and blocks cooperation. Suppose the rich agent transfers an amount $2\,\delta$ to the poor agent (any transfer in a suitable interval works; the value $2\,\delta$ is chosen for simplicity). After the transfer, the new endowments are $\widetilde{e_1} = e_1 + 2\delta, \widetilde{e_2} = e_2 - 2\delta$. The new inequality satisfies $\widetilde{e_2} < (5/4)\widetilde{e_1}$, and, by Proposition 1, the poor agent now strictly gains from switching to b-MKE. The rich agent, although losing part of his endowment, also gains compared to Nash: decreasing $e_2$ reduces his utility in b-MKE continuously, and for sufficiently small delta his utility remains higher than in Nash. Thus, a voluntary transfer from the rich to the poor constitutes a Pareto improvement over Nash and removes the inequality barrier. Determining the exact conditions under which redistribution is guaranteed to work is a subject for future research.

### 4.2. Which Parametrization to Choose?

Proposition 2 demonstrates that the choice of "scale" itself is a source of conflict. The strategic non-equivalence of MKE, noted by Roemer [2019] and Sher [2020], acquires a new dimension here: agents have opposite preferences regarding the space—contributions or consumption—in which to apply Kantian optimization. As shown above, this conflict is not superficial: it reflects a fundamental difference between two understandings of Kantian fairness—solidarity in burden (b-MKE) and solidarity in outcome (c-MKE). In the former case, agents universalize sacrifice; in the latter, the right to private consumption. Both logics are internally consistent and appeal to symmetry, but they lead to different distributive consequences.

More generally, as shown by Sher [2020] and endorsed by Roemer [2020], the problem of defining the "sameness" of actions is fundamental to the Kantian approach. Our model confirms this insight: different notions of what it means to act "in the same way" lead to different distributive outcomes. The decision of which parametrization to adopt becomes a strategic move that determines the distribution of payoffs and requires either a focal point or an explicit coordination procedure.

### 4.3. The Equilibrium-Selection Problem: Efficiency without Boundaries

Proposition 3 shows that in c-MKE the equilibrium set forms a continuum of Pareto-efficient feasible outcomes. A related institutional-design result suggests that different points on the frontier can be supported as Kantian equilibria under suitable rules [Sloev, Lianos, 2026]. Our model, however, highlights the flip side of this possibility: it generates an equilibrium-selection problem that is analogous to the distributional indeterminacy highlighted by the Second Welfare Theorem. Kantian optimization guarantees efficiency but offers no answer as to *which* efficiency should be chosen. This turns cooperation into a bargaining game in which the disagreement point (e.g., the Nash equilibrium) plays a key role. Which conception of justice—utilitarianism, Rawlsianism, the Nash bargaining solution—would be implemented in the absence of an explicit coordination mechanism remains an open question.

### 4.4. Dynamic Stability: A Conceptual Reconstruction

Building on the formal results obtained above, one can offer a conceptual reconstruction of the interaction as a sequence of three steps. The following scheme is not a formal model; it is a way of structuring the insights derived from the analysis:

*Step 1 (cooperation decision):* agents decide whether to adopt the Kantian norm or remain at Nash (unanimity required).

*Step 2 (choice of "scale"):* if cooperation is accepted, agents choose between b-MKE and c-MKE.

*Step 3 (equilibrium selection):* if c-MKE is selected, agents bargain over a concrete point on the frontier; failure to agree results in reversion to Nash.

One possible interpretation of the preceding results is that a chain of expectations may arise. Expectations about the third step (the ability to agree on a point) may influence the choice of parametrization at the second step. In turn, expectations about the second step (which scale will be chosen) may determine the poor agent's willingness to enter cooperation at the first step. In such a chain, the static entry threshold $e_2 < (5/4)e_1$ ceases to be a fixed number and becomes a function of the society's institutional capacity to coordinate the subsequent steps (see, e.g., Skarzhinskaya and Tsurikov, 2017a, 2017b, for related models of coordination and efficiency in collective action). A consequence of this interdependence may be an "institutional trap": agents voluntarily select a coarser scale (b-MKE) in order to avoid the risk of disagreement at the third step, even if c-MKE potentially allows a fairer distribution. Thus, institutions that ensure coordination at the third step may be a condition not only for fairness but also for the very stability of Kantian cooperation.

### 4.5. Research Program

The proposed conceptual scheme allows us to formulate a number of questions for further research that flow directly from the model and are amenable to formal analysis.

- *Inequality (Proposition 1):* Proposition 1 suggests a general question: how does the threshold depend on the number of agents and on the distribution of endowments? Which pre-cooperation redistribution mechanisms can remove the inequality barrier without creating distortionary incentives? How do expectations about future steps formally affect the entry threshold?

- *Conflict of parametrizations (Proposition 2):* Proposition 2 raises the question of whether it is possible to design a parametrization under which MKE automatically coincides with one of the classical conceptions of justice (e.g., the Nash bargaining solution). What determines a society's choice of a particular "scale" when there are more than two agents?
- *Equilibrium selection (Proposition 3):* Proposition 3 leads to the question of which procedures (voting, arbitration, proportional rules) can ensure successful agreement on a point of the Pareto frontier within a formal model, and how they affect the stability of cooperation.

## 5. Conclusion

Using a simple model of a public good with heterogeneous agents, this paper has demonstrated the logical sequence of three challenges that Kantian cooperation faces under inequality. The model serves three interrelated purposes. First, it provides a concrete example that confirms the existence of each of the three effects: the threshold condition at which inequality blocks cooperation and the possibility of overcoming it through voluntary redistribution; the conflict of interest arising from the choice of parametrization; and the equilibrium-selection problem that emerges from a continuum of efficient equilibria. Second, direct extensions of the model make it possible to pose a series of formal questions that constitute a program for further research. Third, the proposed three-step conceptual reconstruction takes the analysis beyond the model and shows how these barriers are linked by expectations about future decisions, thereby opening new perspectives for studying the stability of Kantian cooperation.